\documentclass[runningheads]{svmult}

\usepackage{makeidx}   
\usepackage{graphicx}  
\usepackage{subeqnar}  
\usepackage{multicol}  
\usepackage{physprbb}  
\makeindex             

\begin{document}
\title*{Nineteenth and Twentieth Century Clouds Over 
\newline the Twenty-First Century Virtual Observatory}

\toctitle{Nineteenth and Twentieth Century Clouds Over 
\protect\newline \newline the Twenty-First Century Virtual Observatory}
\titlerunning{Clouds over the Virtual Observatory}

\author{P. J. E. Peebles}

\authorrunning{P. J. E. Peebles}

\institute{Joseph Henry Laboratories, Jadwin Hall\\ Princeton
University, Princeton NJ 08544 USA}

\maketitle              

\begin{abstract}
Physical science has changed in the century
since Lord Kelvin's celebrated essay on {\it Nineteenth Century
Clouds over the Dynamical Theory of Heat and Light}, but some
things are the same. Analogs in what was happening in
physics then and what is happening in astronomy today serve to
remind us why we can be confident the Virtual Observatory of the 
twenty-first century will have a rich list of challenges to
explore. 
\end{abstract}

\section{Introduction}

Astronomy has enjoyed a very good century. Have the basic
problems now been solved, leaving for the astronomers of the
21$^{\rm st}$ century the task of working out the pesky details?  
The question is little discussed -- astronomers are too busy
with ongoing research -- but worth considering from time to time.
I shall argue that we have a useful guide to the long-term
prospects for research in astronomy from analogs to the present
situation in was happening in physics 100 years ago. In both
cases there is a basis of fundamental concepts that are
strikingly successful, apart from some stubborn clouds, or, as we
would now say, challenges for research. 
The clouds over electromagnetism and thermal physics at the start
of the 20$^{\rm th}$ century foreshadowed relativity and quantum 
physics. We can't say what will be learned from the clouds over
present-day astronomy -- I shall mention aspects of the dark
sector, strong space curvature, and the meaning of life --  but
we can be sure they will continue to drive difficult but
fascinating research in astronomy for quite some time to come.   

\section{Physics at the Start of the 20$^{\rm th}$ Century}

The elements of the situation in physics a century ago have been
retold to generations of students, and rightly so; these are 
golden moments in the history of physical science. And I think
they are an edifying example for our assessment of the present
state of research in astronomy.   

At the start of the 20$^{\rm th}$ century physicists had good
reason to believe they had securely established laws of
electromagnetism and thermal physics, well tested in the
laboratory  and applied in rapidly growing power and
communications industries; Lord Kelvin's fortune came from his
contributions to the design of the transatlantic telegraph cable. 
But he and others were well aware of flaws, or clouds, in the 
physics, as famously summarized in Kelvin's~\cite{kelvinpeebles}
essay in 1901.  

Kelvin's Cloud No. I is the luminiferous ether. The experimental
situation is also discussed in Lecture VIII, \emph{The Ether}, in
Michelson's \emph{Light Waves and Their
Uses}~\cite{michelsonpeebles}. You can read about the familiar
experiments -- Michelson's discovery of the isotropy of the
velocity of light, and the Fizeau measurement of the addition of
velocities of light and fluid in a moving fluid -- and others
that are less celebrated but as remarkable. My favorite among the
latter is the measurement of annual aberration in a telescope
that is filled with water so as to reduce the velocity of light.
The results are no surprise to us, but a real problem for Kelvin
and Michelson. Kelvin mentions with approval the contraction idea
of  Fitzgerald and Lorentz, but concludes ``I am afraid we must
still  regard Cloud No. I as very dense''~\cite{kelvinpeebles}.
Einstein's brilliant insight cleared the cloud, and gave us
special relativity theory. If that had not happened I have to
believe people would soon have pieced together the full theory
from these remarkable measurements.     

Kelvin's Cloud II is the inconsistency of the law of partition
of energy at thermal equilibrium with the measured ratios
$C_p/C_v$ of heat capacities of gases at constant pressure and
volume.\footnote{This memorable story has been taught to
generations of students in the introduction to quantum  
mechanics, some of whom I hope actually appreciated it.
Equipartition in classical mechanics says that at thermal
equilibrium at temperature $T$ the mean energy belonging 
to each quadratic term in the Lagrangian is $kT/2$. It follows
that if each atom or molecule in a gas has $\nu$ quadratic terms
associated with its internal structure then, taking account of
the $pdV$ work at constant $p$, the ratio of heat capacities is  
$C_p/C_v = (5+\nu )/(3+\nu )$. Thus classical physics predicts
that a gas of point-like particles has $C_p/C_v = 5/3$, and a gas
of atoms with a rich internal structure, so $\nu$ is large, has
$C_p/C_v$ close to unity.} Kelvin~\cite{kelvinpeebles} quotes
Rayleigh's assessment~\cite{rayleighpeebles}: ``The difficulties
connected with the application of the law of equal 
partition of energy to actual gases have long been felt. In the
case of argon and helium and mercury vapour the ratio of specific
heats (1.67) limits the degrees of freedom of each molecule to
the three required for translatory motion. The value (1.4)
applicable to the principal diatomic gases gives room for three
kinds of translation and for two kinds of rotation. Nothing is left
for rotation round the line joining the atoms, nor for relative
motion of the atoms in this line. Even if we regard the atoms as
mere points, whose rotation means nothing, there must still exist 
energy of the last-mentioned kind, and its amount (according to
the law) should not be inferior.'' Something certainly is wrong.
Kelvin accepted the mechanics and questioned the assumption of
strict statistical equilibrium. Planck (1900) hit on the fix, to
the mechanics, in the model for blackbody radiation, and Einstein
(1907) applied the fix to heat capacities, in early steps to
quantum physics.   

It is often said, at least in introductory remarks in courses on
modern physics, that people were a lot more impressed by the
successes of physics in 1900 than by the clouds, that the feeling
was that physics is essentially complete, apart from fixing a 
few problems and adding decimal places. The famous example
is Michelson's statement (in~\cite{michelsonpeebles},~p.23), that the
``more important fundamental laws and facts of physical science
have all been discovered, and these are now so firmly established
that the possibility of their ever being supplanted in
consequence of new discoveries is exceedingly remote.'' This is
clear enough, and Badash~\cite{badashpeebles} shows Michelson
repeated these sentiments elsewhere, so at the time he must have
meant it. But consider Michelson's summary statement in the same
book, at the end of the chapter on the ether
(\cite{michelsonpeebles},~p.~163): ``The phenomenon of the
aberration of the fixed stars can be accounted 
for on the hypothesis that the ether does not partake of the
Earth's motion in its revolution about the sun. All experiments
for testing this hypothesis have, however, given negative results,
so that the theory may still be said to be in an unsatisfactory
condition.'' And earlier in the summary he says ``Little as we
know about it [the ether], we may say that our ignorance of
ordinary matter is still greater.'' Here Michelson sounds
like someone who sees very real challenges.

These challenges drove hard work, as in Fizeau's remarkable
waterworks and Michelson's~\cite{michelsonpeebles} massive
arrangements to suppress vibrations: ``the apparatus was mounted
on a stone support, about 
four feet square and one foot thick, and this stone was mounted
on a circular disc of wood which floated in a tank of mercury.''
I see no evidence of complacency in Kelvin's~(\cite{kelvinpeebles},
p. 17) struggle to visualize mercury vapor atoms, which are
capable of producing a rich line spectrum but at thermal
equilibrium in laboratory conditions seem to be incapable even of
rotating, or in Rowland's 1899 presidential address to the    
American Physical Society~\cite{rowlandpeebles}:  
``What is matter; what is gravitation; what is ether and the
radiation through it; what is electricity and magnetism; how are
these connected together and what is their relation to heat?
These are the greater problems of the universe. But many
infinitely smaller problems we must attack and solve before we
can even guess at the solution of the greater ones.''    

Badash~\cite{badashpeebles} gives a valuable survey of opinions
across a broader range of the academic community, and concludes
that at the end of the 19$^{\rm th}$ century the idea that
science is reaching completeness `was more a 
``low-grade infection,'' \emph{but nevertheless very real}.' This
sounds right, but my impression is that the infection had little
effect on the research of leading physicists, including
Michelson.

The confidence in the established parts of physics at the start
of the 20$^{\rm th}$ century was well placed: we still use and
teach this electromagnetism and thermodynamics -- though we now
think of it as part of a hierarchy of approximations that for
all we will ever know may run arbitrarily deep. Concerns about
the 19$^{\rm th}$ clouds could not have anticipated the vast
enlargement of physics and our worldview in the 20$^{\rm th}$
century, but the point for our purpose is that the clouds were
recognized and driving research.    

I offer some parallels to the situation in present-day
astronomy. We know how stars like the Sun shine, but there are 
big gaps in our understanding of how stars form, at high redshift 
and even in our own galaxy. I classify star formation as a
Rowland-type ``smaller problem:'' it is fiendishly difficult but
approachable by well-motivated lines of research involving
standard physics (as far as we know). Such Rowland-type problems
are the key to a  
healthy science, and astronomy has them in abundance. We know the  
universe is evolving, and the evidence is that general relativity 
theory gives a good description of the dynamics. But we don't
know what the universe is made of -- apart from the five percent
or so in the visible sector -- or what happens when spacetime 
curvature gets large, as was the case in the very early universe
and happens now in the centers of galaxies. These are
Kelvin-level clouds: critical issues whose resolution would
greatly advance our understanding of the material world. We don't
know what the present-day clouds are hiding, but we can be sure
they will continue to be a good focus for research.   

\section{Astronomy at the Start of the 21$^{\rm st}$ Century}

The situation in astronomy in 1900 was close to the academic myth
about physics. Badash~\cite{badashpeebles} gives a good quote
from Newcomb~\cite{newcombpeebles}: ``we do appear 
to be fast approaching the limits of our knowledge $\ldots$ one
comet is so much like another that we cannot regard one as adding
in any important degree to our knowledge. The result is that the
work which really occupies the attention of the astronomer is
less the discovery of new things than the elaboration of those
already known, and the entire systemization of our knowledge.'' 
The main systemization was the cataloging of angular positions, 
apparent magnitudes, and spectral classifications of literally 
hundreds of thousands of stars. But this dreary labor led to
wonderful new things; consider these two examples of research 
trajectories.\footnote{I have 
taken the liberty of indicating contributions by several people,
and even groups, under the name of a representative leading
figure, with an approximate year for developments that in some
cases occurred over many years. I hope it is understood that
another reviewer could choose very different representative 
examples of what happened in 20$^{\rm th}$ century astronomy.
Harwit~\cite{harwitpeebles} presents a well-documented and much
more complete analysis of 
discoveries in astronomy and the prospects for discoveries of
new astronomical phenomena.}

Eddington's (1924) gas spheres gave Bethe (1938) the physical
conditions for nuclear reactions in stars, and a way out of  
the discrepancy between the Helmholz--Kelvin (1860) Solar
cooling time and the much greater geological times from
radioactive decay ages. A beautiful recent development is the  
demonstration that the Solar neutrino luminosity really is 
in satisfactory agreement with the theory of the Solar nuclear
reaction rates, to be understood with the help of the
demonstration of nonzero neutrino masses. 

Kapteyn (1901) set the distance scale for star counts in our
island universe, Shapley (1918) enlarged the island, and Hubble
(1925) placed it in the near 
homogeneous realm of the nebulae. Hubble's linear relation
between his distances to the nebulae and Slipher's (1914)
redshifts led Lema\^\i tre (1927) to the now standard model for
the expanding universe. The most direct evidence that our 
universe actually is evolving -- expanding and cooling -- was
completed with the demonstration by the USA COBE and Canadian UBC
experiments (1990) that the 3~K 
cosmic background radiation spectrum is very close to thermal. In
the 1930s Hubble commenced the great program of cosmological
tests to check the relativistic Friedmann--Lema\^\i tre
model for the expanding universe. Now, seven decades
later, we are approaching a satisfactory application of the
tests, which the relativistic cosmology passes so far. 

A byproduct of the cosmological tests is evidence
that structure grew out of a mass distribution at high
redshift that is specified by one function of one variable, the
near scale-invariant power spectrum of a random Gaussian process.
There are problems with details, as will be discussed, but the
evidence pretty strongly indicates this is a good approximation
to the way it is. The Rowland-type problem, of breathtaking
scope and complexity, is to demonstrate that standard physics
actually can account for the origin of the worlds and their 
spectacular variety of phenomena out of this simple initial
condition.  

I have mentioned stories with some happy endings, in reasonably
conclusive resolutions of lines of research that have occupied
generations of astronomers. We cannot say whether more happy
endings to big puzzles are in store, but we get some
feeling for the prospects by considering present-day clouds over
astronomy. I shall comment on two from the 20$^{\rm th}$ century
and one from the 19$^{\rm th}$ century.     

\subsection{Cloud No. I: the Dark Sector}

The dark sector includes the nonbaryonic matter that is thought
to dominate the outer parts of galaxies and clusters of galaxies;
Einstein's cosmological constant, $\Lambda$, or dark energy that
acts like it; and the vacuum energy density. The darkest
part of the cloud is over the vacuum energy. I draw these
comments from a review of the issues in~\cite{rppeebles} and the
executive summary in~\cite{peebpeebles}.  

Nernst~\cite{nernstpeebles} seems to be the first to have discussed the
energy of the quantum vacuum, in 1916. His zero-point energy for
each mode of 
oscillation of the electromagnetic field is off by a factor of
two, remarkably good considering this was before Heisenberg and
Schr\"odinger. Nernst showed that the sum over zero-point
energies of the modes with laboratory wavelengths is on the order
of 1~g~cm$^{-2}$. Pauli (in~\cite{paulipeebles},~p. 250) was quite 
aware that this mass density would be ruinous for relativistic
cosmology; he advised that we just ignore the zero-point energy
of the electromagnetic field. This is a prescription, of course,
and not even a rational one. 
Pauli certainly knew that one must take account of zero-point
energies to get the right binding energies in nonrelativistic
particle mechanics. We now know the same applies to gravitational
masses. And in standard physics the zero-point energies of fields
are just as real. The problem with the vacuum energy density has 
persisted -- if anything grown more puzzling -- through all the
spectacular advances in physics in 
the 20$^{\rm th}$ century. I like Wilczek's phrase: this aspect
of our physics is ``profoundly incomplete''~\cite{wilczekpeebles}. It
is a Kelvin-level cloud: within physics that is wonderfully well 
tested and successful in a broad range of applications there is a
distinct glitch.

We have observational probes that might be helpful. 
If the vacuum presents the same properties to any inertial
observer, its effect on spacetime curvature is the same as
Einstein's cosmological constant, $\Lambda$. The evidence from the
cosmological tests is that the expansion of the universe actually
is dominated a term that acts like $\Lambda$ -- though the
absolute value is ridiculously small compared to what is
suggested by current ideas in particle physics. The case for
detection is 
serious, but since it depends on difficult observations and
insecure models I am inclined to limit the odds to maybe five to
one. But work in progress should convincingly show us whether
a term that acts like $\Lambda$ really is present. 

Until recently the tendency in the astronomy community has been
to hope that it could get by with Pauli's prescription, or at
worse the phenomenological description of the vacuum by the
numerical value of one constant, $\Lambda$, leaving the
dispersal of this cloud to the physicists. But current ideas are
that $\Lambda$ is only an approximation to a dynamical entity,
dark energy, whose mass density varies with time on the scale of
cosmic evolution, and varies with position in response to the
large-scale irregularities in the matter distribution. Detection
of these effects would not solve the vacuum energy density
problem, but it would be a spectacularly stimulating clue. We
know how 
it might be done, and I have been hearing ambitious plans to make
the astronomical measurements. You may be sure the physicists
will be hanging on every word of progress; they are desperate for
something to knock them off dead center.

In the standard cosmology the dark sector also contains
nonbaryonic matter that dominates the mass in the outer dark
halos of galaxies and the mass in clusters of galaxies. I am in
sympathy with those who ask for more evidence this nonbaryonic
matter really exists, but I think the case already is close to
compelling. The clearest exhibition of dark matter is the giant
luminous arcs -- the gravitationally lensed images of background
galaxies produced by the gravitational deflection of light by the
masses in clusters of galaxies. No force law I can imagine could
produce these smooth arcs out of gravitating matter with the
clumpy distribution of the starlight in clusters. There has to be 
cluster dark matter, and if it were baryonic it would cause ugly
problems~\cite{rppeebles}. 

We have little empirical guidance to the physics of the dark
sector: we are working in the dark. We accordingly
adopt the simplest physics we can get away with, which is good
strategy, but certainly need not be the whole story: consider 
that polytropic ideal gas spheres were good enough for
Eddington's analysis of the structure of the Sun, but
helioseismology reveals a host of new details. If our model for
the dark sector is missing details that matter it will be
revealed by problems in fitting the observations. And there are
hints of problems, from observations of the structure and
formation of galaxies. My list is headed by the prediction that
elliptical galaxies form by mergers at modest redshifts, which
seems to be at odds with the observation of massive quasars at 
$z\sim 6$; the prediction of appreciable debris in the voids
defined by $L_\ast$ galaxies, which seems to be at odds with the
observation that dwarf, irregular, and $L_\ast$ galaxies share 
quite similar distributions; and the prediction of cusp-like dark
matter cores in low surface brightness galaxies, which is at odds 
with what is observed. These are Rowland-type problems that draw
on the rich phenomenology of astronomy, from the latest 
observations by the Hubble Space Telescope to the vast
accumulation of lore from decades past. Sorting through all this
takes time, but I expect will show us whether the problems with
the standard picture for the dark sector will be resolved by
better understanding of the observations and theory, or will be
promoted to a Kelvin-level cloud. 

\subsection{Cloud No. II: Strong Spacetime Curvature}

Cloud II is the singularities of general relativity, where the 
theory becomes meaningless. It took some time for people to sort
out the physical singularities from singular
coordinate labels, and to face up to the phenomenological 
importance of the former. I remember as a graduate student in the
late 1950s reading a distinguished physicist's elegant picture of
the bounce in an oscillating universe: like turning a glove
inside out, one 
finger at a time. In the mid 1960s Penrose's~\cite{penrosepeebles}
pioneering approach to singularity theorems forced us to accept
that we need deeper physics to see past the formal singularity at
infinite redshift in the relativistic Friedmann--Lema\^\i tre 
cosmological model. At about the same time, the discovery of 
quasars, and the  broader recognition of active galactic nuclei,
offered an example of strong spacetime curvature in compact
objects closer to hand. Now, a half century later, we have 
rich phenomenologies of compact objects and cosmology, and we
still have the singularities.

Analyses of the astrophysics of massive compact objects -- those
observed at 
the centers of large galaxies, and star remnants more massive
than a white dwarf -- usually take as given a Schwarzschild or
Kerr black 
hole geometry with a truly black inside, in discussions of what
have grown to be quite detailed observations. There are no problems
with this approach, a sign of the remarkable predictive power of
general relativity theory. But good science demands that we seek
positive evidence in support of the black hole picture, and watch
for credible evidence that the standard picture may not 
be quite right. Maybe advances in fundamental physics will show
us what really is happening in the centers of galaxies, or maybe
the dispersal of this cloud will be guided by the phenomenology.  

Analyses of observations in cosmology finesse the
formal singularity of the Friedmann--Lema\^\i tre
model, and the unknown physics at the Planck scale, by
stipulating initial conditions at a more modest redshift, 
let us say $z=10^{15}$. 
Nowadays the initial conditions often are given a pedigree,
from the inflation model, and the observational constraints on
the initial conditions are used to infer conditions on what was
happening during inflation. But, since the inflation scenario can
fit a considerable range and variety of initial conditions, we
don't know whether these measures of the very early universe
amount to anything more than a ``just so'' story. Three
assignments may help. 

We look to observational astronomers and cosmologists for tighter 
constraints on the initial condition at redshift $z=10^{15}$.
And it behooves us to watch for hints that there is more 
to learn about cosmic evolution than is encoded in this initial  
condition within the present standard cosmology. The successes
of the extrapolation of standard physics to the length and time 
scales of cosmology are impressive, but the enormous
extrapolation certainly allows room for surprises. I am watching
for them in the problems with galaxy formation I mentioned in
connection with the dark sector. 

We look to those exploring ideas about the early universe to try
to find alternatives to inflation. If all due diligence yielded 
none we would have an argument by default that inflation really  
happened, a dismal closure but better than nothing. Alternatives
are under discussion; it will be of great interest to know
whether some variant of the ekpyrotic
universe~\cite{ekpyroticpeebles}
has a physical basis comparable to that of inflation, which is
not asking all that much.

We look to the physics community to build a firmer basis for
cosmology at high redshift. If fundamental physics converged on a
complete theory that 
predicts a definite version of inflation, or some other picture
for the early universe, which agrees with the astronomical 
constraints, it will convincingly complete cosmology. The  
prediction's the thing, of course.  

\subsection{Cloud No. III: the Meaning of Life}

This is a cloud over a much broader community. We
can leave to the experts in other fields the philosophical
issues, and the analysis of the molecular basis for life. The
task for astronomy and its Virtual Observatory is to search for
evidence of extraterrestrial life. This is a  Kelvin-level cloud:
a powerful driver of research whose outcome could profoundly
affect our worldview. 

Maybe life on Earth came from primitive extraterrestrial seeds;
Hoyle and Wickramasinghe~\cite{hoylepeebles} survey the history
and present state of ideas. Maybe there are advanced forms of
life on other worlds, seeded or evolved out of spontaneously
created life. The familiar 19$^{\rm th}$ century example of the 
search for organized life 
is Lowell's study of possible signs on Mars; the search continues
in the SETI and OSETI projects. The Terrestrial
Planet Finder (TPF) will search for Earth-like worlds where life
might flourish in a primitive or organized state. 

I read that the search for extraterrestrial life is the part of  
astronomy that most interests most people. I offer four
observations of how the big ideas and activities in society have
influenced the directions of this research.

First, Charles Darwin's deeply influential arguments for
evolution by natural selection forced debate on what the first
step in the evolution of life might have been. At about the same
time, people were coming to the conclusion that spontaneous
generation is an exceedingly rare event, if it happens at all, and
maybe contrary to Darwin's principle that life evolves out
of life~\cite{farleypeebles},~\cite{crowepeebles}. It was natural  
therefore that people turned to the idea of extraterrestrial
seeds. Helmholtz (1874), a most influential physicist and
physiologist, argued for the idea, as did an 
important chemist, Arrhenius (1908). Kelvin (1871) endorsed the 
general idea, but not natural selection: he argued for
``intelligent and benevolent design''~\cite{kelvin2peebles}.

Second, the end of the 19$^{\rm th}$ 
century was a time of large-scale civil engineering, including
completion of the modern Suez Canal in 1869. It is perhaps not so  
surprising that Lowell looked for signs of big engineering on
Mars.

Third, this is an age of computers and information transfer. 
I think it's not surprising that people are searching for
extraterrestrial bar codes. I don't mean to mock serious and
important science: a source of bar codes 
would signify self-aware life by any definition. Imagine the
effect on our society of the demonstration that there actually is 
extraterrestrial self-aware life, that might even have something 
to say to us. 

Fourth, this is an age of big science, that is supported by the
wealth of nations. A logical consequence is that research in
science is influenced by big government. The TPF is a recent
example: this is pure curiosity-driven big science that 
originated within government funding agencies, rather than being
forced by intense pressure from a scientific community. 

I offer two lessons from these observations. First, the
fascination with the idea of life on other worlds has a long
history, back through the 19$^{\rm th}$ century, and, I expect,
it has a long future. But societies evolve, and it is natural to 
expect the focus of the search for extraterrestrial life will
evolve too.  

Second, the means of support of the scientific enterprise are
evolving; the TPF is leading the curve. The TPF certainly may
yield wonderful results;  
we have the inspiring precedent of Slipher's discovery of the
cosmological redshift, at the observatory Lowell built with a
goal paralleling that of the TPF. But there is the difference
that funding agencies have to tend to many masters; they can't
have the compulsive attention span of curiosity-driven 
people like Lowell. The Virtual Observatory is not leading this
curve: a community is fighting for it, in the style of what gave
us the space telescope, and what happened in physics in the last
half century. These are generally happy examples -- apart from
such glitches as sunset clauses -- of what I suppose is an
inevitable  development: the directions of research in astronomy
are increasingly influenced by government as well as society, and 
astronomers must continue learning how to deal with it.

\section{Concluding Remarks}

Our ability to explore the physical universe is limited by
resources and intellectual energy: the scientific enterprise must
eventually reach completion by exhaustion. But we can be sure
this will not happen any time soon to astronomy and its Virtual
Observatory, because the subject has a rich list of Rowland-type  
problems to address, and, as I have discussed, a key role to play
in the exploration of clear and present Kelvin-level gaps in our
understanding of the fundamental basis for physical 
science. There was no guarantee in 1900 that the clouds over
physics would clear, with a wonderful expansion of our knowledge. 
It would be foolish to try to guess what the present clouds
might foreshadow, but we can list the general possibilities. 
Maybe the clouds will resist all efforts at resolution. If so,
convincing people of this certainly will generate a lot of work
for astronomers. Maybe the clouds will be 
cleared and at last leave astronomers to tidy up the pesky
details. Or maybe clearing the clouds will reveal a new set, as
has happened before. 

I have avoided until now commenting on a serious issue under
debate in the astronomy community: is this an appropriate
time to commit limited resources to an International Virtual
Observatory? I respect the arguments against, but am persuaded by
personal experience that 
the growth of the Virtual Observatory is inevitable and would
benefit from intelligent design. Two years ago the walls of 
my office were covered by about 25 meters of journal rows, 
dating back to 1965. I loved the convenience of reaching for a
copy of the wanted article. But I've discarded the journals; I
love even more the much greater convenience and power of ADS,
arXiv, and JSTOR. I notice many colleagues feel the same: we have
become addicted to these Virtual Libraries. Present-day Virtual
Observatories are a useful but limited counterpart. Their
further development seems to me to be an inevitable part of what
we see happening around us, and surely calls for the
proactive community response I have observed at this meeting.   

\section*{Acknowledgments}
I have benefitted from advice from Larry Badash, Neta Bahcall,
Jeremy Bernstein, Masataka Fukugita, Rich Gott, Martin Harwit,
Gerald Holton, Stacey McGaugh, Bharat Ratra, Paul Schechter, Max
Tegmark, and Ed Turner. This work was supported in part by the 
USA National Science Foundation.

\end{document}